# Anomalous Superconducting Proximity Effect in Hybrid Oxide Heterostructures with Antiferromagnetic Layer

G. A. Ovsyannikov[1,2], P. Komissinskiy[1,2,3], I. V. Borisenko[1], Yu. V. Kislinskii[1], A. V. Zaitsev[1], K. Y. Constantinian[1], and D. Winkler[2].

[1]Institute of Radio Engineering and Electronics Russian Academy of Sciences, Moscow, Russia
e-mail: gena@hitech.cplire.ru
[2]Department of Microtechnology and Nanoscience, Chalmers University of Technology, Gothenburg, Sweden
[3]Department of Material Sciences, Darmstadt University of Technology, Darmstadt, Germany

Abstract

We report an investigation of current transport in $Nb/Au/Ca_{1-x}Sr_xCuO_2/YBa_2Cu_3O_{7-\delta}$ heterostructures including an antiferromagnetic interlayer ($Ca_{1-x}Sr_xCuO_2$). Epitaxial thin films of $YBa_2Cu_3O_{7-\delta}$ and $Ca_{1-x}Sr_xCuO_2$ were grown by laser ablation on $NdGaO_3$ substrates; the thicknesses of $Ca_{1-x}Sr_xCuO_2$ films were 20 and 50 nm, and x=0.15 or 0.5. Our experimental results show that the superconducting current transport is increased in the $Nb/Au/Ca_{1-x}Sr_xCuO_2/YBa_2Cu_3O_{7-\delta}$ heterostructures in comparison with the ones observed in previously studied $Nb/Au/YBa_2Cu_3O_{7-\delta}$ heterostructures. It was found that the critical current of the structure is very sensitive to the absolute value and direction of the external magnetic field.

Unconventional properties of superconducting hybrid junctions having interlayers comprised by magnetic materials are of great interest for fundamental physical studies and electronic applications. Studies of Josephson junctions with metallic ferromagnetic interlayers have been made [1-6]. There is an opportunity to manipulate properties of junctions with an antiferromagnetic (AF) interlayer by weak external magnetic fields.

The idea was suggested by Gorkov and Kresin in [5], where they predicted an anomalously strong magnetic field dependence of the superconducting critical current of Superconductor-Antiferromagnet-Superconductor (S-AF-S) Josephson junctions. However, a strong magnetic field dependence of the superconducting critical current was not observed in the case of polycrystalline thin film S-AF-S (Nb-FeMn-Nb) junctions [6]. The giant proximity effect, namely, an anomalously large value of the superconducting critical current has been found in cuprate oxide structures with thick AF oxide inter-layers [7, 8]. An alternative explanation of the giant proximity effect has been suggested in [9], where the experimentally obtained large value of the superconducting current is explained by a number of micro-shorts in the interlayer. Therefore, a clarification is needed.

Here we report on our studies of the current transport in hybrid thin film Mesa HeteroStructures (MHS) of $Nb/Au/Ca_{1-x}Sr_xCuO_2/YBa_2Cu_3O_x$, where Nb is a conventional s-wave superconductor (S), $YBa_2Cu_3O_x$ (YBCO) is an oxide superconductor having a dominant d-wave symmetrical component of the superconducting order parameter (D), and $Ca_{1-x}Sr_xCuO_2$ (CSCO) is a G-type AF, where quasiparticle spin direction alternates in the {111} ferromagnetic planes. We use epitaxial AF thin films, where the magnetization direction is rigidly locked to the orientation of the crystallographic axes. Taking into account weak magnetic coupling between the neighbour magnetic planes we can consider current transport along CSCO [111] being equivalent to the case of canted AF.

Double-layer epitaxial thin film heterostructures CSCO/YBCO were grown *in-situ* by pulsed laser deposition on (110) and (320)$NdGaO_3$ (NGO) substrates. Thus, the c-axis of the YBCO/CSCO heterostructures are perpendicular to the substrate surface or tilted from the substrate normal by the angle $\gamma=11°$. Typically the d=20 or 50 nm thick $Ca_{1-x}Sr_xCuO_2$ (CSCO) films were deposited on top of the 150 nm thick YBCO films. The Sr doping of the CSCO films at the levels of x=0.15 and x=0.5 was realized by varying the target composition. The obtained YBCO/CSCO heterostructures were later on covered by 10-20nm thick Au films and 200nm thick Nb films deposited by DC-magnetron sputtering in Ar atmosphere. In order to fabricate Nb/Au/CSCO/YBCO MHS we applied conventional photolithography, reactive plasma etching and Ar ion-milling techniques. A $SiO_2$ layer was deposited by RF-magnetron sputtering and patterned afterwards in order to define the area of the MHS. An additional 200 nm thick Nb films was deposited on top of the mesa area and patterned in order to form the superconducting wiring. Thus, Nb/Au/CSCO/YBCO MHS with areas from 10×10 up to 50×50 $\mu m^2$ have been successfully fabricated [10].

The values of the peak-to-valley surface roughness of single-layer (001)YBCO and (001)CSCO thin films measured by atomic force microscopy are about 2 nm and increase up to 5 nm for the CSCO/YBCO heterostructure grown on (110)NGO substrate. The growth of the YBCO thin film on the (320)NGO results in 20 nm high surface growth steps, which are tilted from the substrate surface by an angle of about 11°. This is approximately equal to the tilting of the c-axis of the YBCO (and CSCO) film with respect to the substrate normal direction.

Our earlier studies showed hopping conductivity of CSCO thin films at temperatures 4.2-300 K [11]. The Neel temperature of the CSCO (x=0.15) films, $T_N$=90-120 K, was measured by an electronic paramagnetic resonance technique [12]. The obtained values of the $T_N$ differ from earlier neutron diffraction experiments on powder samples, possibly, because of difference in oxygen content between the two cases [13].

A 4-point measurement technique was used for electrical characterization of our Nb/Au/CSCO/YBCO MHS: two contacts to the YBCO electrode and two contacts to the Nb counter electrode. At temperatures below the superconducting critical temperature of the YBCO film, $T_c$, the wiring, in fact, is identical to the classical 4-point measurement scheme. At $T>T_c$ the resistance of the YBCO wiring should be taken into account. Thus, at $T>T_c$ the R(T) dependence of the MHS is similar to a single YBCO film due to the large resistance of the YBCO film (Fig. 1). At $T_c'<T<T_c$, where $T_c'$ is the critical temperature of the Nb film, the MHS resistance is almost independent on temperature and determined by the interface resistances between the materials CSCO/YBCO, Au/CSCO, and Nb/Au, plus the resistance of the CSCO interlayer. Taking into account the independently measured resistance of the CSCO film $\rho \cong 1$ $\Omega$ cm, $R_{CSCO}$~0.1 $\Omega$ [11, 12] and the Nb/Au interface characteristic resistance ~$10^{-12}$ $\Omega \cdot cm^2$ [10], $R_{Nb/Au}$~1 $\mu\Omega$, we conclude that their contribution to the 5 $\Omega$ resistance of the MHS is small. Assuming a high quality of the CSCO/YBCO interface (which is the case for our in-situ epitaxial thin film deposition process) the $R_{CSCO/YBCO}$ should also be negligibly small because of the close values of the Fermi velocities for the materials. Thus, the resistance of the Au/CSCO interface determines the total resistance of the MHS. Note that the values of the Au/CSCO interface resistance vary with the substrate tilt angle $\gamma$ and Sr doping x, but are of the same order of magnitude as the ones of the Au/YBCO interface. Totally we have not found any significant difference in DC and RF properties of the MHS at different values of x and angle $\gamma$.

Regarding the MHS resistance at $T_c'<T<T_c$ we may have two competing quasiparticle transfer channels acting in opposite ways on the R(T): tunneling through the potential barrier at the Au/CSCO interface [14] and Andreev bound states at the CSCO/YBCO interface [15]. The Andreev bound states manifest themselves in a conductance peak observed in I-V curves at low bias voltages, V<5 mV [15] (Fig.2). Similar conductance peaks have been observed earlier, for example, in the I-V curves of the Nb/Au/YBCO MHS on tilted NGO substrates [10]. In general, the I-V curves of our Nb/Au/CSCO/YBCO MHS are of the hyperbolic shape at V≤ 1 mV, which is typical for Josephson junctions. No YBCO gap features have been observed in the I-V curves. However, at $T<T_c'$ the features at V≤ 2 mV associated with the superconducting gap of Nb are present (Fig.2).

The inset of Fig. 1 shows the temperature dependence of the critical current $I_c(T)$ of the Nb/Au/CSCO/YBCO MHS. $I_c(T)$ dependence of the Nb/Au/YBCO MHS (d=0) is presented for comparison. In spite of the 50nm thick CSCO interlayer in the MHS we observe a good qualitative agreement of both $I_c(T)$ dependences. The $I_c(T)$ curves follow the temperature dependence of the Nb superconducting gap (filled squares in the inset of Fig. 1) [16,17]. However, we have observed neither a dependence of the $I_cR_N$-product with the thickness of the CSCO layer nor a square-law increase of the critical current with decreasing temperature. Thus, we may conclude that the coherence length, $\xi_{AF}$, which determines the penetration depth of the superconducting ordering in the AF layer, is not small with respect to the thickness of the CSCO layer d. Otherwise an exponential decrease of the superconducting critical current with increasing temperature should occur in our Nb/Au/CSCO/YBCO MHS [18].

A qualitative estimation of the $\xi_{AF}$ value can be made by modelling the AF as a stack of -F-F'-F-F'- thin (comparable to $\xi_{AF}$) ferromagnetic (F, F') layers with alternating direction of the exchange field parallel to the plane of the layers. We can analyse the coordinate dependences of the superconducting pairing functions $f_\pm = f_\pm (x, \omega_n, \cos\theta)$ (corresponding to opposite orientations of the electron spin, $\omega_n = \pi T(2n+1)$ is the Matsubara frequency) using the Eilenberger equation (see, for example, [1-3]). Our analysis shows that the values of the penetration length of the superconducting pairing function into the -F-F'-F-F'- stack are the same as in the case of a homogeneous exchange field [2]. Therefore we may conclude that the condition $\xi_{AF} > d$ may be realized only in the case of ballistic electron transport through the AF layer, i.e. the long mean free path in the AF $l > d$. Taking into account $H_{ex}\tau/\hbar \gg 1$, where $\tau = l/v_F$ one can obtain that $\xi_{AF} \sim \min\{l, \hbar v_F/(kT)\}$, that coincides with the case of the homogeneous exchange field [2]. Thus, the value of $\xi_{AF}$ may not be small with respect to d.

Assuming that the quasiparticle mean free path $l \sim 10$ nm in the CSCO film (approximately equal to the thickness of the interlayer) at 4.2 K, we evaluate the value of the Fermi momentum $p_F = \hbar k_F$ by using the condition $k_F l \sim 1$ [14]. The latter estimation presumes the degree of uncertainty of the quasiparticle Fermi velocity value in the CSCO film $v_F$. Assuming that the effective mass is equal to the electron mass, we obtain $\xi_{AF} \sim \hbar v_F/(6\pi kT) \approx 7$ nm at T=4.2 K. Thus, the interlayer thickness d is comparable with $\xi_{AF}$.

Our experimental data show that the values of the $I_cR_N$-product of the Nb/Au/CSCO/YBCO MHS are 2-3 times higher than those of the Nb/Au/YBCO MHS (see Table 1). Note that we also present the data obtained for the MHS with the c-oriented YBCO in the table 1. Thus, one can conclude that the reason of the observed enhancement of the superconducting current transport in the MHS is related to the presence of the CSCO interlayer. At the moment we may only suggest possible reasons of the observed enhancement of the superconducting current transport in the Nb/Au/CSCO/YBCO MHS.

One reason may be related to an increase of the s-wave component of the order parameter at the Au/CSCO interface with respect to the s-wave component of the order parameter at the Au/YBCO interface. It is well-known that the order parameter of YBCO has dominant d-wave ($\Delta_d$) and sub-dominant s-wave ($\Delta_s$) components. In the Nb/Au/YBCO MHS in c-axis oriented YBCO films and small transparency of the Au/YBCO interface, the non-vanishing value of the superconducting current is due to a non-zero value of $\Delta_s$ [17]. A similar conclusion can be made for the Nb/Au/CSCO/YBCO MHS with the c-oriented YBCO and small transparency of the Au/CSCO interface. In this case a non-vanishing value of the superconducting current is due to a non-zero value of the s-wave order parameter in CSCO at the Au/CSCO interface, $\Delta'_s$. Therefore, one may conclude that the observed enhancement of the superconducting current transport may be related to the enhancement of $\Delta'_s$ with respect to the value of $\Delta_s$. Note, that similar enhancement of the s-wave component of the superconducting order parameter at the HTS/normal metal interface was analyzed in [19, 20].

The calibration of the CSCO thin film deposition rate and thickness was performed by measuring the thickness of a patterned, 300 nm thick CSCO film deposited on the $NdGaO_3$ substrate. However, the CSCO film used in our Nb/Au/CSCO/YBCO MHS as the AF barrier was grown on top of the YBCO film. Due to the surface roughness of the YBCO template film, the effective thickness of the CSCO film on top of the YBCO may be smaller. The latter statement may be indirectly confirmed by our analysis of the X-ray diffraction data, which correlate with the effective thin film thickness. From Fig. 3 we can see that the values of the Full Width at Half Maximum (FWHM) of the θ-2θ diffraction peaks of the CSCO films on $NdGaO_3$ substrates (0.23°) are smaller than ones of the CSCO deposited on the top of YBCO (FWHM=0.43°) despite the CSCO film on top of YBCO is two times thicker. The latter FWHM value is roughly the same as the one of the YBCO film. At approximately identical thin film structures, the FWHM values may be proportional to the difference in film thickness. In addition the FWHM values of the rocking curve peaks of the CSCO film deposited on top of the YBCO films increased by several times. Thus, the effective thickness of the CSCO film in our MHS may be smaller than the one determined from the deposition process.

Applied monochromatic mm-waves induce Shapiro steps in the I-V curves of all MHS (Fig.4). Oscillations of the Shapiro step amplitude *vs.* applied microwave power (Fig. 5) confirm the Josephson origin of the superconducting current. Moreover, a very good correspondence between the static (obtained from the DC I-V curve measurement) and dynamic parameters of the MHS is observed. The critical frequency $f_c=2eV_c/h=71$ GHz calculated from $V_c=I_cR_N=147$ μV is comparable with $f_c=56$ GHz determined from the maximum of the first Shapiro step within the Resistive Shunted Junction model [21]. The deviation becomes smaller if we take into account a presence of a second harmonic component in the superconducting Current-Phase Relation (CPR), which is clearly indicated in the I-V curve by fractional Shapiro steps (Fig.4). The experimental data in Fig. 5 are well fitted by a theory [22] that takes into account the amplitude, $I_{c2}$, of the second harmonic of the CPR $q=I_{c2}/I_c=0.2$. The result clearly indicates the absence of micro-shorts [9].

Note that a giant Josephson effect caused by an anomalously large penetration of the superconducting order parameter into the ferromagnetic layer was recently observed in an S-F-S structure with singlet s-wave superconductors [23]. The effect was explained by a triplet superconducting pairing, which presumably is generated at the interface of the s-wave superconductor and the ferromagnetic film with spiral type of magnetization. As theoretically predicted in [3, 24], the triplet superconducting pairing may penetrate into the ferromagnetic material (making use of diffusive electron transport) by a much longer distance than the singlet superconducting pairing does. We would like to emphasize that in our case, unlike [24], the value of the Josephson superconducting current may be determined by a singlet component of the superconducting pairing.

Fig. 6 shows the magnetic field dependence $I_c(H)$ of two (20x20 $\mu m^2$) MHS: Nb/Au/YBCO and Nb/Au/CSCO/YBCO. $I_c(H)$ of the Nb/Au/YBCO MHS has a classical Fraunhofer shape presuming a homogeneous distribution of the superconducting current across the MHS area. The first minimum of the $I_c(H)$ dependence corresponds to a flux quantum penetrating into the Nb/Au/YBCO MHS at $\mu_0H_0\approx\pm230$ $\mu T$, which is more than 10 times higher than in the case of the Nb/Au/CSCO/YBCO MHS, $\mu_0H_0\approx\pm5$ $\mu T$. This agrees with a theoretical analysis by Gor'kov and Kresin [5], where an anomalously high magnetic field sensitivity of the Josephson junction with an antiferromagnetic interlayer was predicted. CSCO has a magnetic sub-lattice, which, in general, does not coincide with the crystallographic $CuO_2$ planes of the material as demonstrated in [23]. In fact, we observed changes in the shape of the $I_c(H)$ dependence with the direction of the applied magnetic field. Moreover, the $I_c(H)$ dependence of the larger MHS (50x50 $\mu m^2$) reveals additional oscillations theoretically predicted in [5] at $H<<H_0$. These oscillations may be explained by interaction of the magnetization of the CSCO film with the applied magnetic field [5].

Thus, we have experimentally demonstrated that in Nb/Au/CSCO/YBCO mesa-heterostructures with an epitaxial CSCO AF interlayer the superconducting current transport can be significantly larger than without it. Transformation of the dominant d-wave symmetrical component of the superconducting order parameter into an s-wave component at the interface with an AF interlayer could be crucial for the observed effect. The condensate wave function may penetrate deeply into the antiferromagnetic layer by distances up to tens of nm. The superconducting critical current of the investigated mesa-heterostructures is highly sensitive to the amplitude and direction of an external magnetic field.

The authors thank V.V. Demidov, F. Lombardi, V.A. Luzanov, and I.M. Kotelyanski for helpful discussions of the results. We gratefully acknowledge the partial support of this work by the Russian Academy of Sciences, Russian Foundation for Basic Research, 04-02-16818a; the grants of the president of the Russian Federation: MK-2654.2005.2, the scientific school grant 7812.2006.2, and the FP6 program of EU. Financial support by the Swedish KAW, KVA, SI and SSF "OXIDE" program has been provided.

Figure Captions

Fig. 1. Typical resistance *vs.* temperature dependence of the Nb/Au/CSCO/YBCO Mesa HeteroStructure (MHS) #269b measured at a bias current of 1 µA. The inset shows the temperature dependences of the critical currents $I_c(T)$ of the Nb/Au/CSCO/YBCO MHS (solid line) and Nb/Au/YBCO MHS along the *c*-axis of YBCO (dot line) [16]. The $I_c(T)$ dependences for every MHS are normalized by their $I_c$ at 4.2 K. The temperature dependence of the superconducting gap of Nb $\Delta_{Nb}(T)$ experimentally determined from the I-V curves of the Nb/Au/YBCO MHS along the *c*-axis of YBCO is shown in the inset for comparison (squares). The values of $\Delta_{Nb}(T)$ have been normalized by the value of $\Delta_{Nb}(4.2K)=1.1$ meV. The temperature in the inset is normalized by the temperature of the superconducting transition of Nb $T_c'=9.2$ K.

Fig. 2. Dependence of the electrical conductance *vs.* bias voltage of the Nb/Au/CSCO/YBCO mesa heterostructure #274b for several temperatures. Gap features caused by the Nb electrode superconducting gap structure are clearly observed at T=4.2 K.

Fig. 3. X-ray θ-2θ scans for an epitaxial (001)CSCO film (d=50 nm) on the (110)NGO substrate (the upper graph) and (001)CSCO/(001)YBCO/(110)NGO thin film multilayer structure (d=100 nm) (the lower graph). The rocking curve of (001)CSCO/(110)NGO bilayer is shown in the inset.

Fig.4. I-V curves of the Nb/Au/CSCO/YBCO mesa heterostructure #269a at different powers ($P_e$) of external radiation with frequency $f_e$=56 GHz. Multiple and fractional Shapiro steps are observed at 4.2 K. A Bessel curve corresponding to the $I_c(P_e)$ dependence within a resistive shunted Josephson junction model is included for comparison.

Fig.5. Dependence of the superconducting critical current $I_c$ (filled circles) and the first Shapiro step $I_1$ (triangles) observed in the I-V curves of the Nb/Au/CSCO/YBCO mesa heterostructure #269a *vs.* the applied microwave power $P_e$ at 4.2 K. The black and red lines correspond to the $I_c(P_e)$ and $I_1(P_e)$ curves numerically calculated from the resistive shunted Josephson junction model taking into account the second harmonic in the current-phase relation (q=0.2) [22]. The dotted line shows the calculated $I_1(P_e)$ dependence for q=0, i.e. a Bessel function.

Fig.6. Magnetic field dependences of the critical current $I_c(H)$ of the Nb/Au/CSCO/YBCO #274a (filled circles) and Nb/Au/YBCO mesa heterostructures (open circles) at 4.2 K: The inset shows a detailed view of the $I_c(H)$ graph of the Nb/Au/CSCO/YBCO mesa heterostructrucure.

| Sample # | $d_{sc}$, nm | $\gamma$, grad | x | A, $\mu m^2$ | $I_c$, $\mu A$ | $j_c$, $A/cm^2$ | $R_N$, $\Omega$ | $V_c$, $\mu V$ |
|---|---|---|---|---|---|---|---|---|
| 273 | 20 | 11 | 0.5 | 20x20 | 890 | $2.2 \cdot 10^2$ | 0.15 | 137 |
| 274a | 50 | 11 | 0.5 | 20x20 | 10 | 2.5 | 20 | 200 |
| 274b | 50 | 11 | 0.5 | 50x50 | 70 | 2.9 | 2.9 | 203 |
| 269a | 50 | 0 | 0.15 | 10x10 | 48 | $4.8 \cdot 10^1$ | 3.0 | 144 |
| 269b | 50 | 0 | 0.15 | 20x20 | 555 | $1.4 \cdot 10^2$ | 0.38 | 211 |
| 125 | 0 | 11 | 0 | 20x20 | 18 | 5.0 | 3.6 | 65 |
| 710 | 0 | 0 | 0 | 15x15 | 15 | 6.7 | 5.1 | 76.5 |

Table 1. Parameters of the investigated Nb/Au/CSCO/YBCO mesa heterostructures at T=4.2 K and H=0. $d_{sc}$ is the thickness of the CSCO interlayer, $\gamma$ is the tilt angle of c-axis of the YBCO film from the normal of the substrate, x characterizes the Sr doping level in $Ca_{1-x}Sr_xCuO_2$. Parameters A, $I_c$, $j_c$, $R_N$, and $V_c=I_cR_N$ are the area, superconducting critical current, critical current density, normal resistance, and characteristic voltage of the mesa heterostructures, correspondingly.

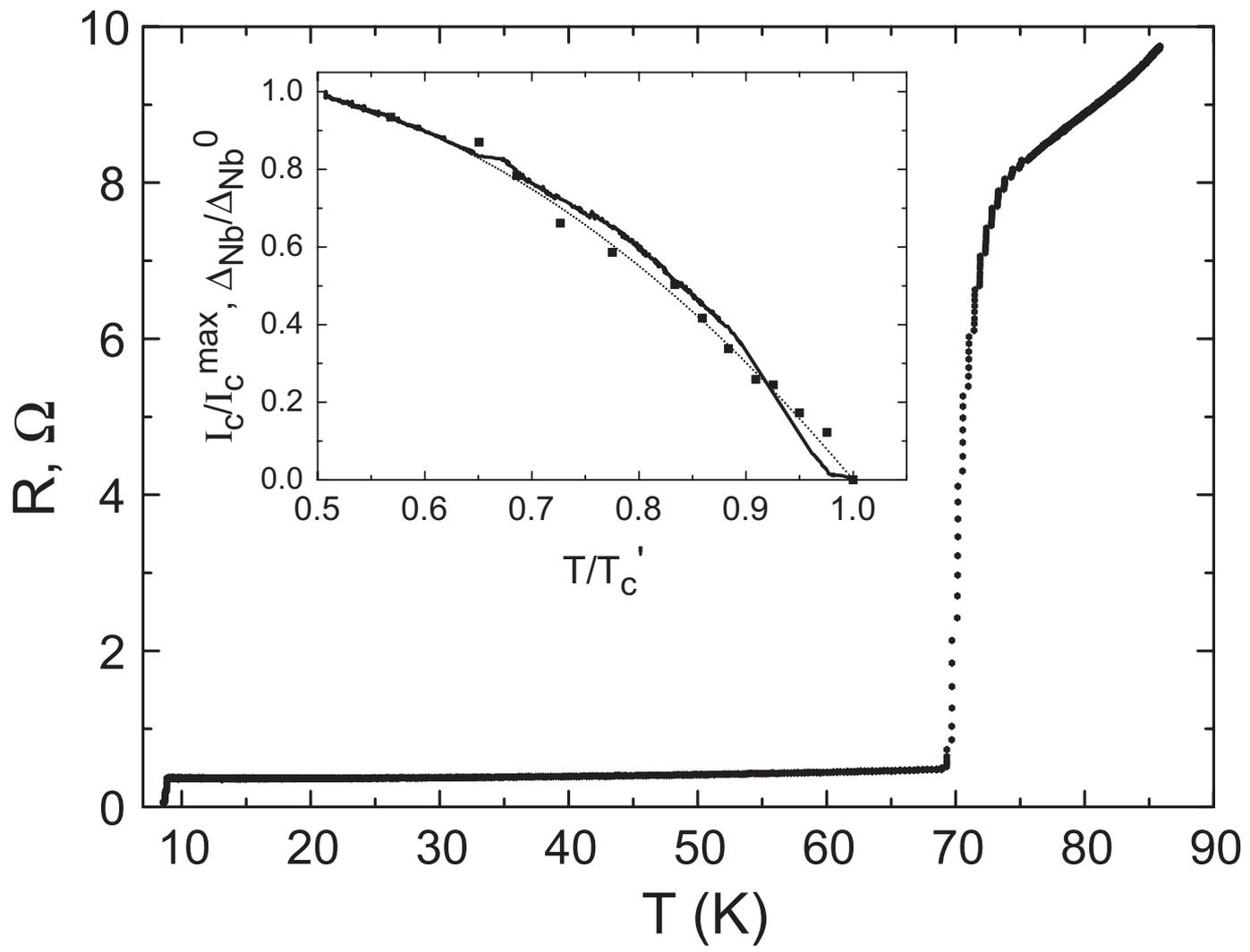

Fig. 1

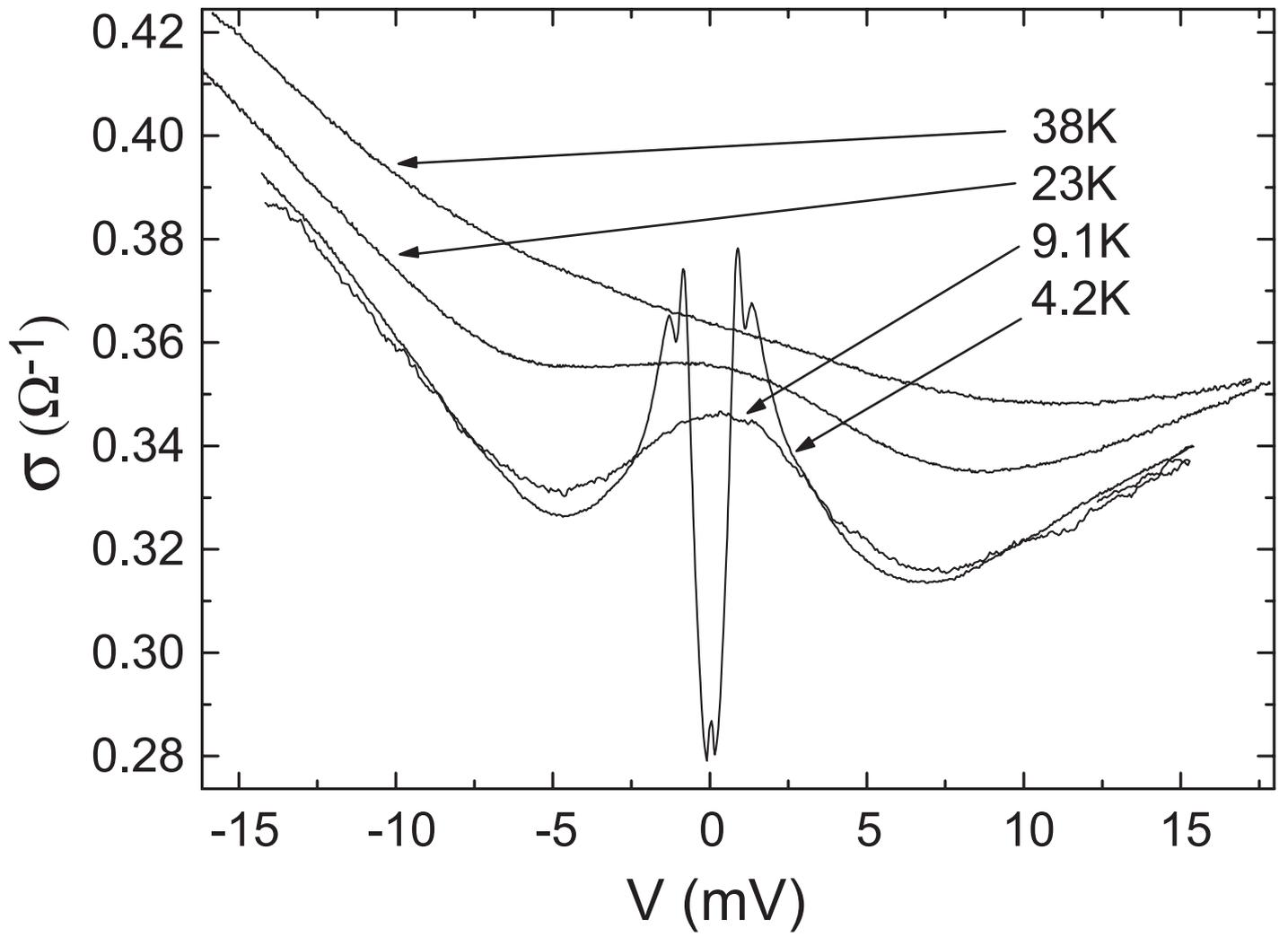

Fig. 2

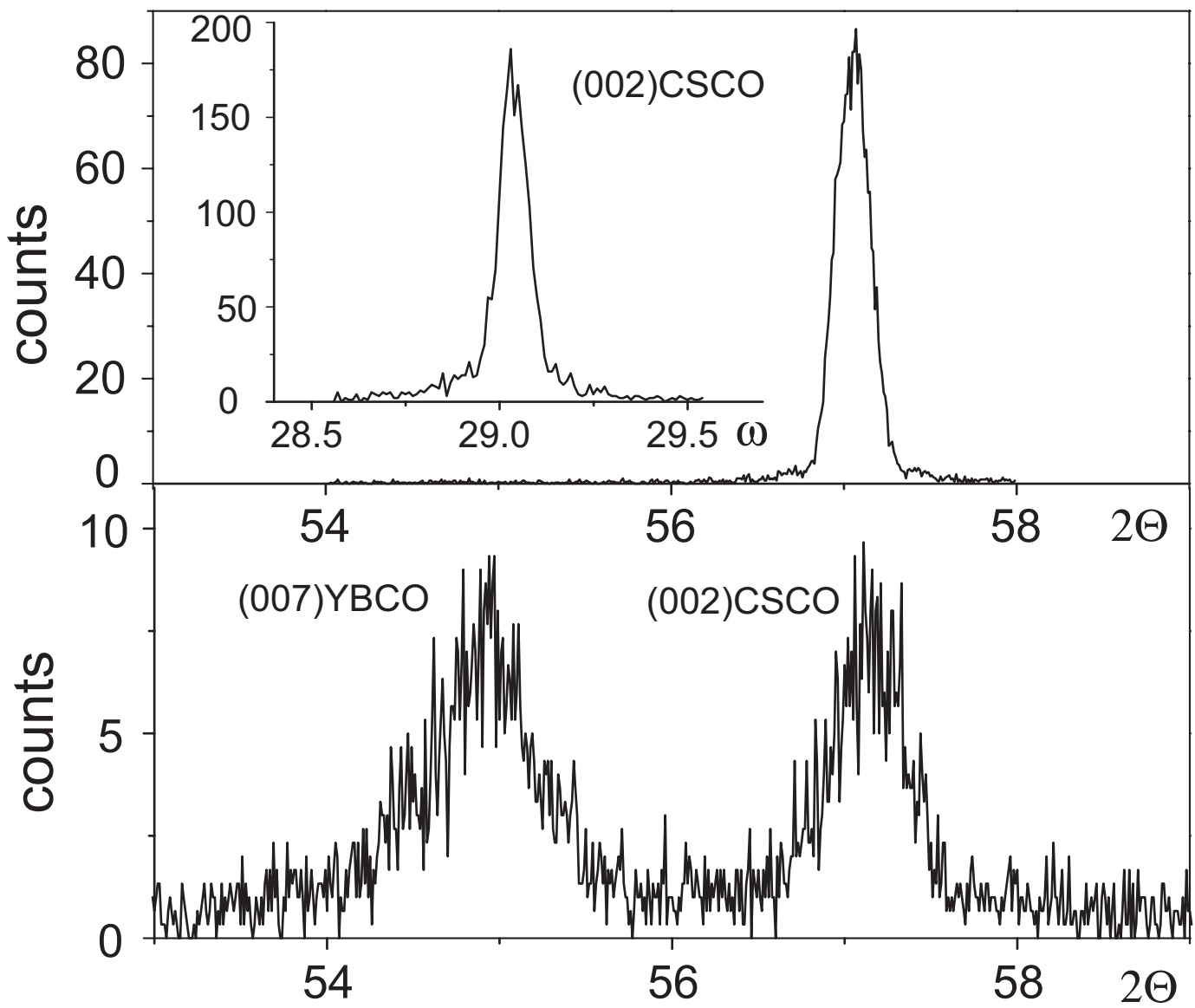

Fig. 3

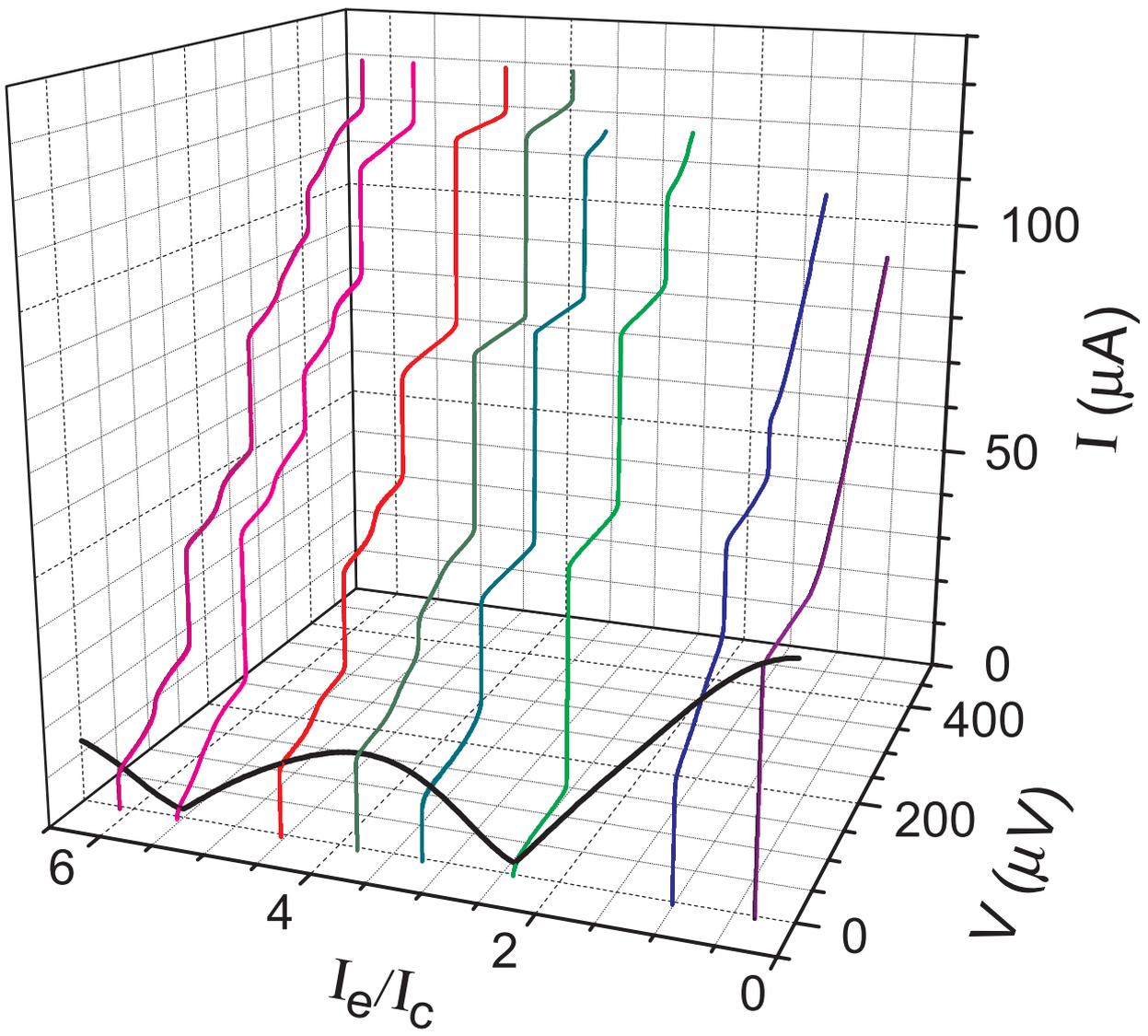

Fig. 4

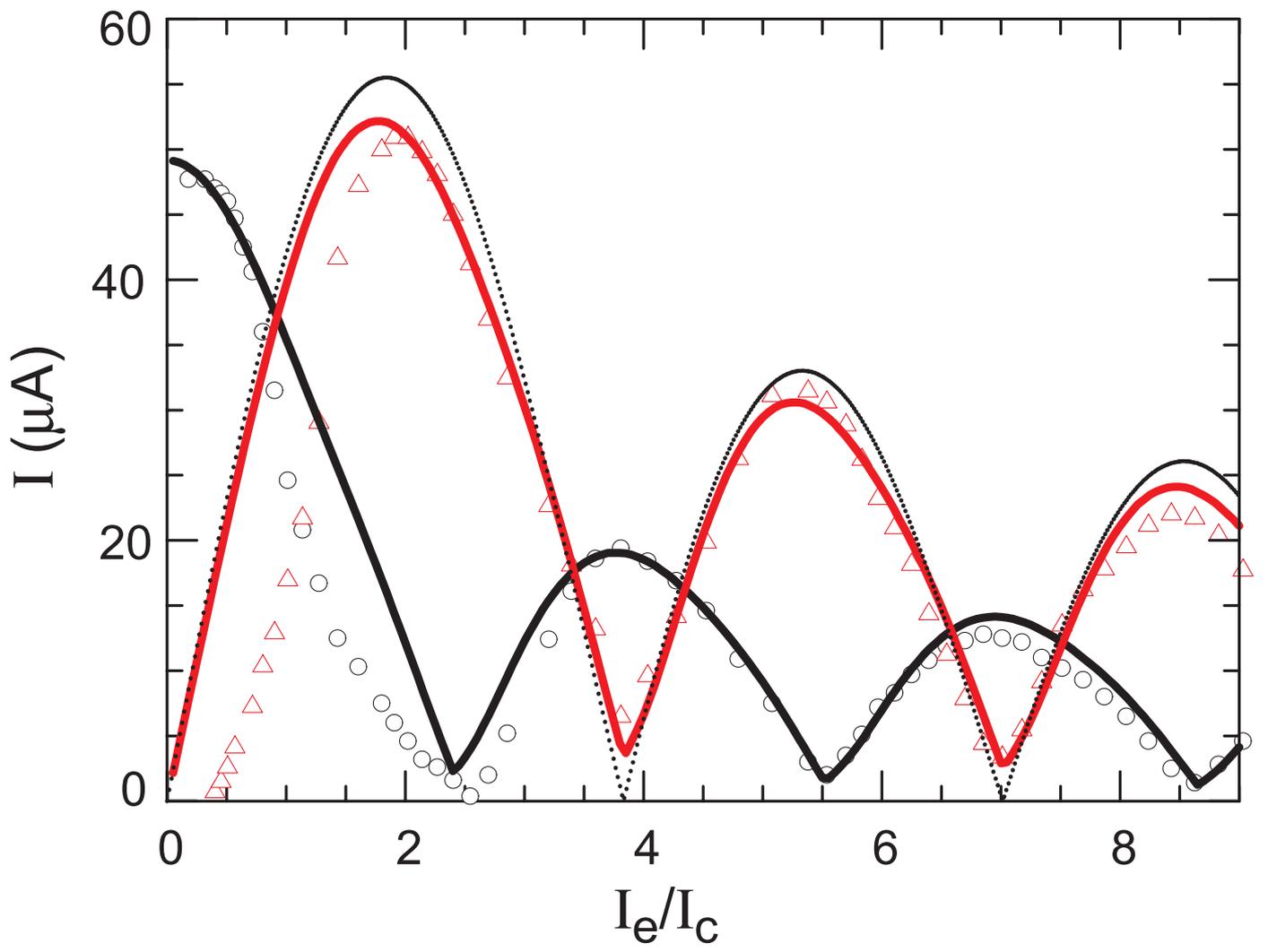

Fig. 5

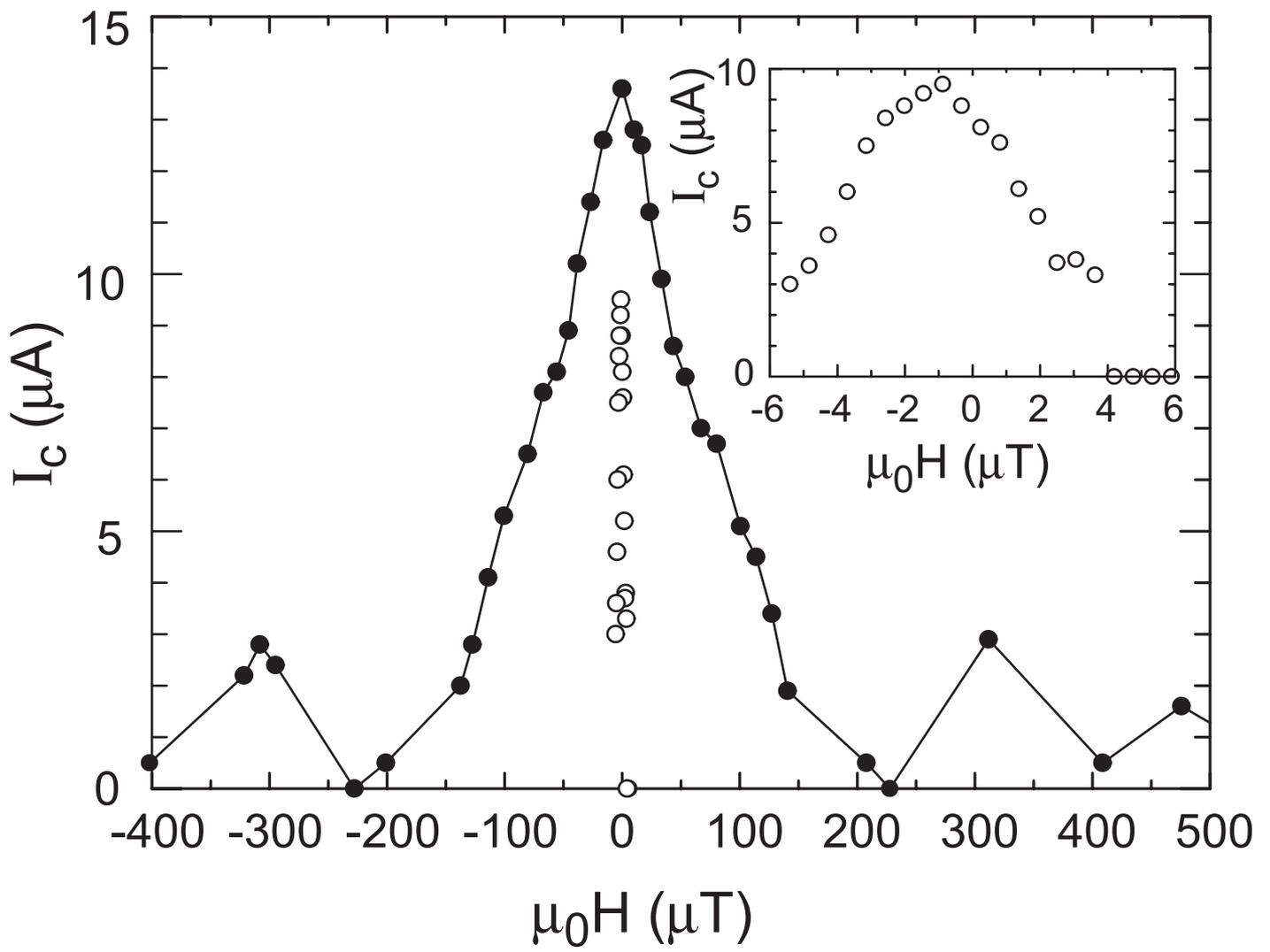

Fig. 6